\documentclass[final,times,3p,twocolumn,number]{elsarticle}
\usepackage{amssymb}
\usepackage{graphicx}
\biboptions{sort&compress}
\journal{Computer Physics Communications}
\begin{document}
\begin{frontmatter}

\title{Time-dependent magnetotransport in semiconductor nanostructures\\
       via the generalized master equation}
\author[hi,ncts]{Vidar Gudmundsson}
\ead{vidar@raunvis.hi.is}
\author[nuu]{Chi-Shung Tang}
\author[hi]{Cosmin Mihai Gainar}
\author[rom]{Valeriu Moldoveanu}
\author[hr]{Andrei Manolescu}
\address[hi]{Science Institute, University of Iceland,
              Dunhaga 3, IS-107 Reykjavik, Iceland}
\address[ncts]{Physics Division, National Center for Theoretical Sciences,\\ 
               P.O.\ Box 2-131, Hsinchu 30013, Taiwan}
\address[nuu]{Department of Mechanical Engineering,
                National United University,\\ 1 Lienda, Miaoli 36003, Taiwan}
\address[rom]{National Institute of Materials Physics, P.O.\ Box MG-7,
              Bucharest-Magurele, Romania}
\address[hr]{Reykjavik University, School of Science and Engineering,\\
             Kringlan 1, IS-103 Reykjavik, Iceland}

\begin{abstract}
Transport of electrons through two-dimensional semiconductor structures on the
nanoscale in the presence of perpendicular magnetic field depends on the interplay
of geometry of the system, the leads, and the magnetic length.
We use a generalized master equation (GME) formalism to describe the transport
through the system without resorting to the Markov approximation.
Coupling to the leads results in elastic and inelastic processes in the system that are
described to a high order by the integro-differential equation of the GME formalism.
Geometrical details of systems and leads leave their fingerprints on the
transport of electrons through them. The GME formalism can be used to describe
both the initial transient regime immediately after the coupling of the leads to the
system and the steady state achieved after a longer time.
\end{abstract}
\end{frontmatter}

\begin{figure}[htbq]
      \begin{center}
      \includegraphics[width=0.42\textwidth,angle=0]{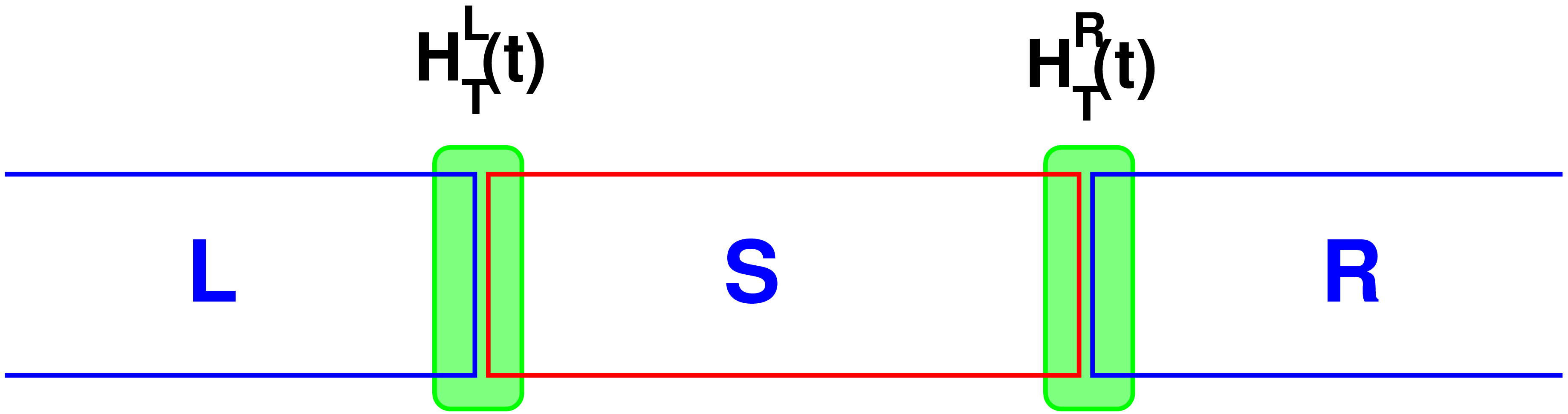}
      \end{center}
      \caption{Schematic of the system, the leads, and the coupling between
               them represented by the tunneling Hamiltonian $H_T^{L,R}(t)$,
               (see \cite{Gudmundsson09:113007})}
      \label{System}
\end{figure}

Commonly, various transport formalisms have been tried on very simple ``model systems''.
Here, we show that the generalized master equation can be used to investigate
magnetotransport properties of a two-dimensional electron system with nontrivial
geometry. Furthermore, we find that the geometrical shape of the leads coupled to the
system strongly influence the transport. The GME formalism has been used by several
groups to study transport \cite{Harbola06:235309,Welack06:044712,Timm08:195416,Vaz08:012012}.
To derive the non-Markovian GME we project the Liouville-von Neumann equation for the
density operator of the leads and the system on the system by tracing out all operators
pertaining to the leads, obtaining the reduced density operator (RDO) $\rho_{\rm S}(t)$
describing the evolution of the system under the influence of the leads (the reservoirs).
The derivation had its origin in quantum
optics \cite{Zwanzig60:1338,Nakajima58:948}. For technical details see \cite{Moldoveanu09:073019}
for a lattice model and \cite{Gudmundsson09:113007} for a continuous model. The semi-infinite
leads are coupled to the finite system at $t=0$. Here, we will assume no Coulomb interaction
between the electrons in the system, but we have added the Coulomb interaction via ``exact
diagonalization'' in a different communication \cite{Moldoveanu10:IGME}. 
An external magnetic field
with strength $B$ is perpendicular to the 2D electron system in the leads and the system.
The derivation of the GME is carried out with the assumption of weak tunneling coupling
between the system and the leads. In the continuous model the coupling is described by a
nonlocal overlap integral between states in the system and the wire in the contact area
marked green in Fig.\ \ref{System}, (see \cite{Gudmundsson09:113007}). This gives us a
complex coupling scheme between the leads and system that depends on the geometry of the
subsystems instead of a single coupling constant often used.

\begin{figure}[htbq]
      \begin{center}
      \includegraphics[width=0.17\textwidth,angle=0]{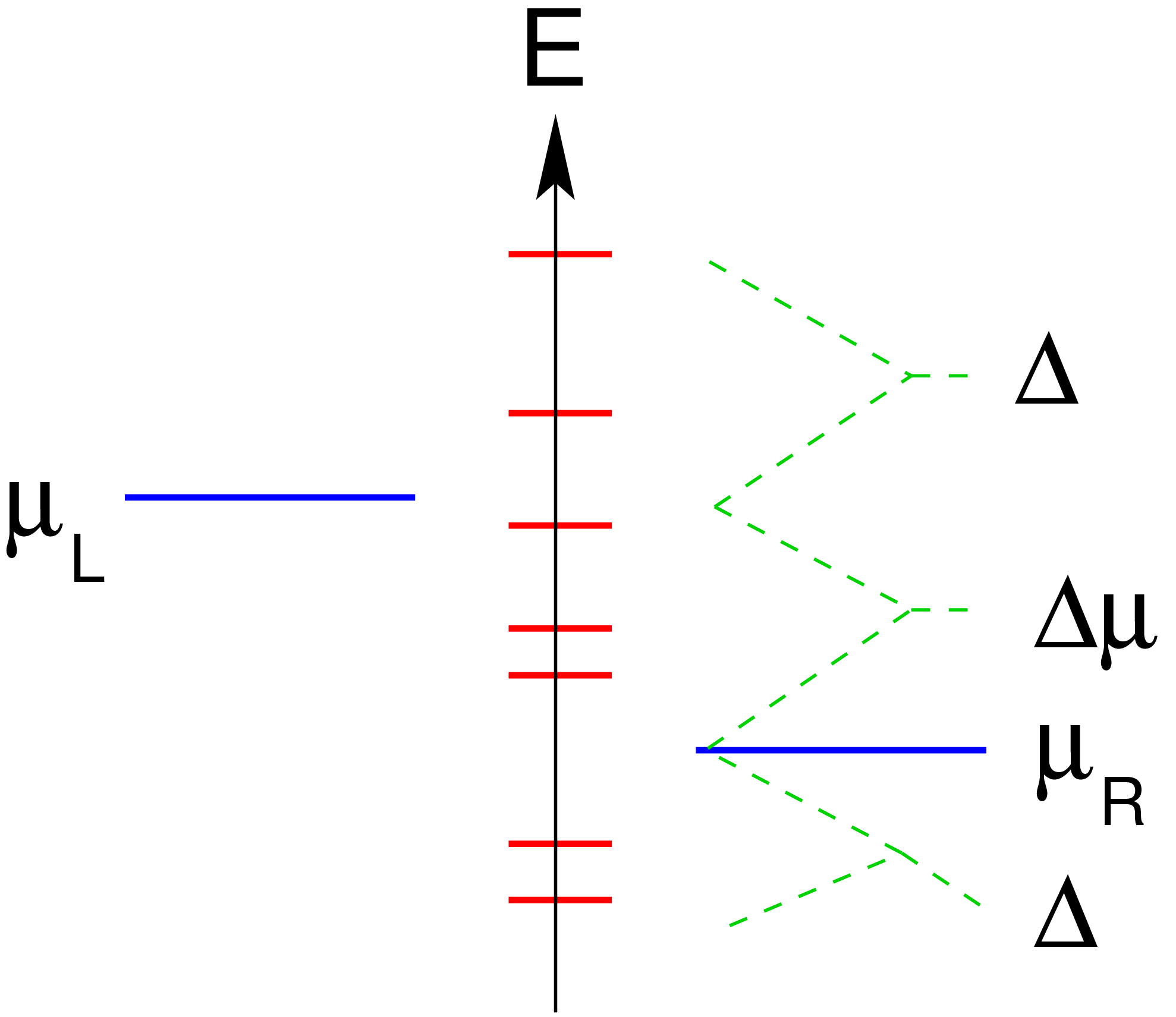}
      \end{center}
      \caption{Schematic of the relevant single electron states (SESs) in and around the bias
               window included in the transport calculation.}
      \label{BiasWindow}
\end{figure}

The GME-formalism is a many-electron formalism and thus we select as relevant for the
transport single-electron states (SESs) in and around the bias-window shown in
Fig.\ \ref{BiasWindow} to build the necessary many-electron states (MESs).
In the calculation here we use 10 SESs, and since it is carried out at
magnetic field $B=1.0$ T we have to remember that some states have the
character of an edge state while others are bulk states. The coupling to the
states will vary with their character, and the magnetic field influences this
coupling to the largest extent.

\begin{figure}[htbq]
      \begin{center}
      \includegraphics[width=0.42\textwidth,angle=0]{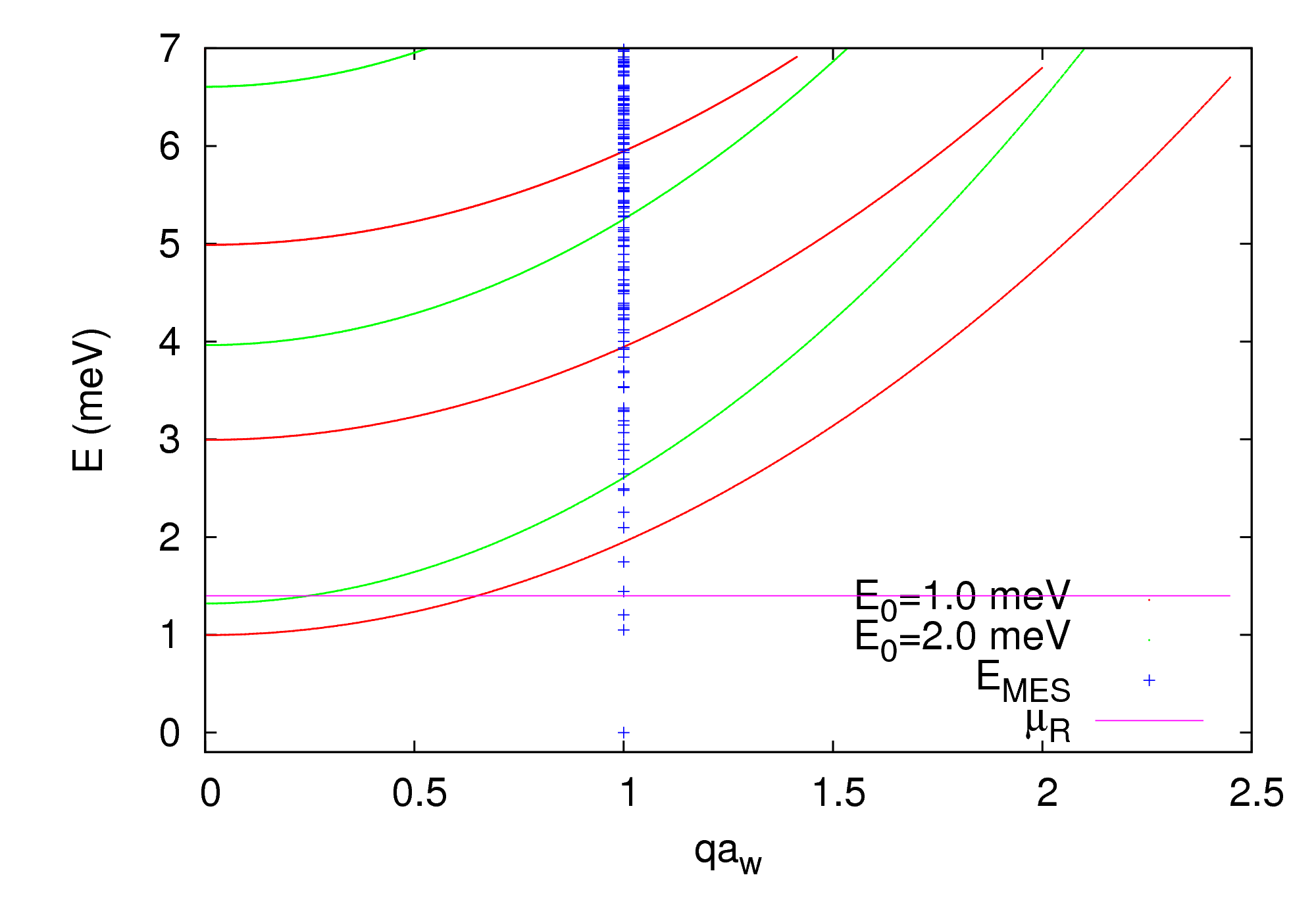}
      \end{center}
      \caption{The discrete spectra of the MESs of the system, the continuous SESs spectra
               for a broad or narrow semi-infinite lead, and the chemical potential in the
               right lead $\mu_R$.}
      \label{Spectra}
\end{figure}

\begin{figure}[htbq]
      \begin{center}
      \includegraphics[width=0.42\textwidth,angle=0]{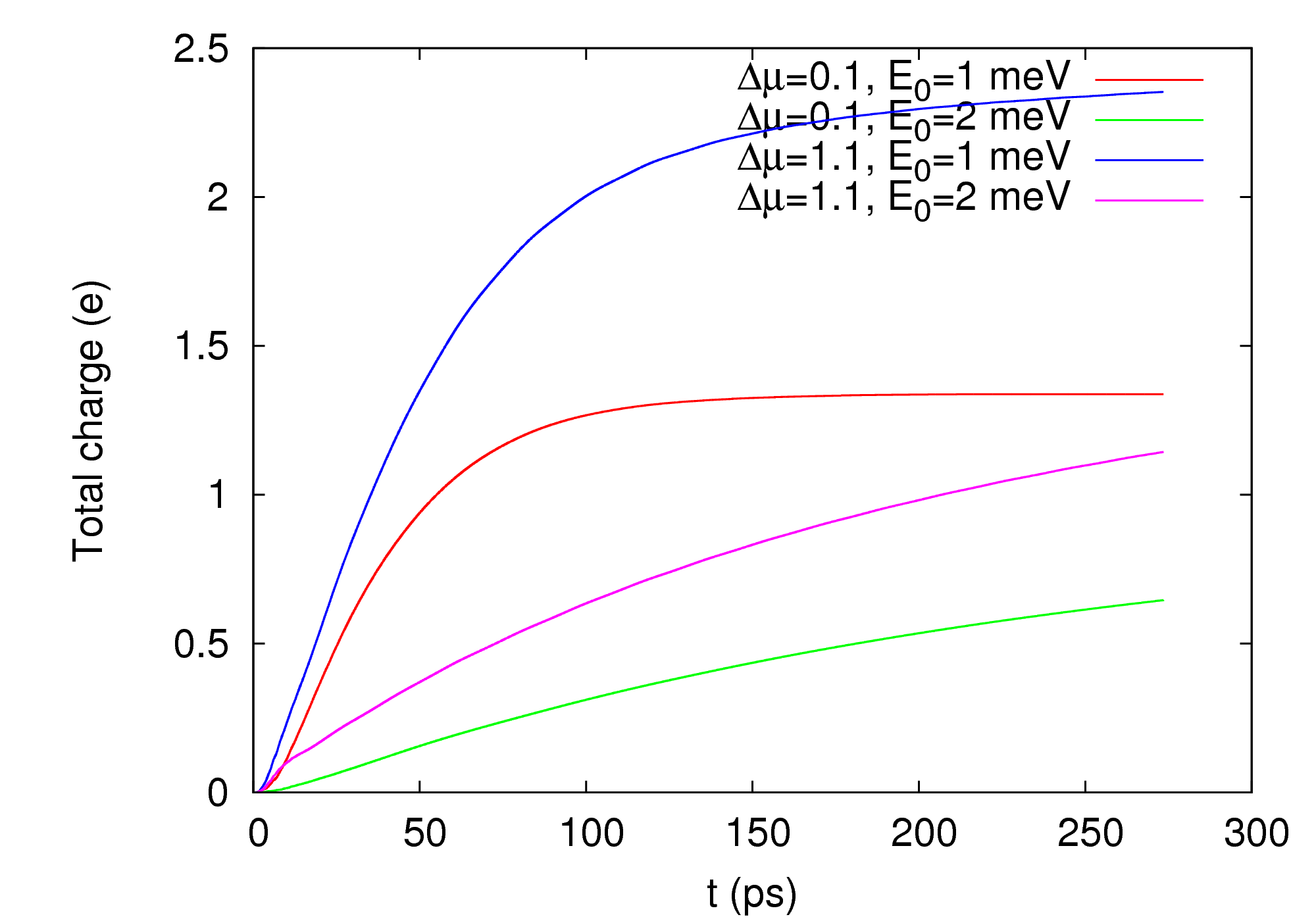}
      \end{center}
      \caption{The total charge in the system as a function of time for different
               values of the bias $\Delta\mu$ and confinement of the leads
               $E_0=\hbar\Omega^{R,L}_0$.}
      \label{TotalCharge}
\end{figure}

\begin{figure}[htbq]
      \begin{center}
      \includegraphics[width=0.42\textwidth,angle=0]{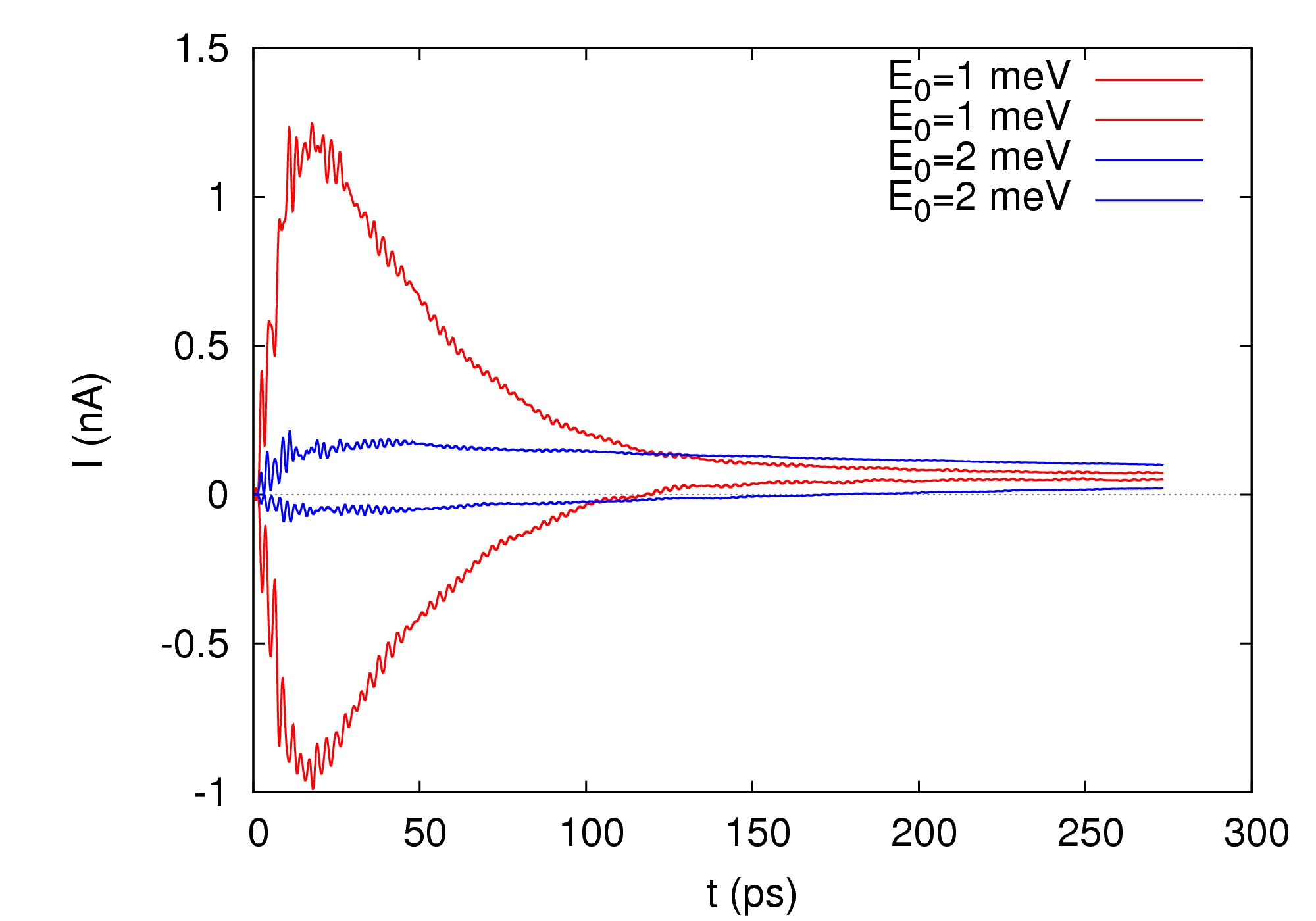}
      \includegraphics[width=0.42\textwidth,angle=0]{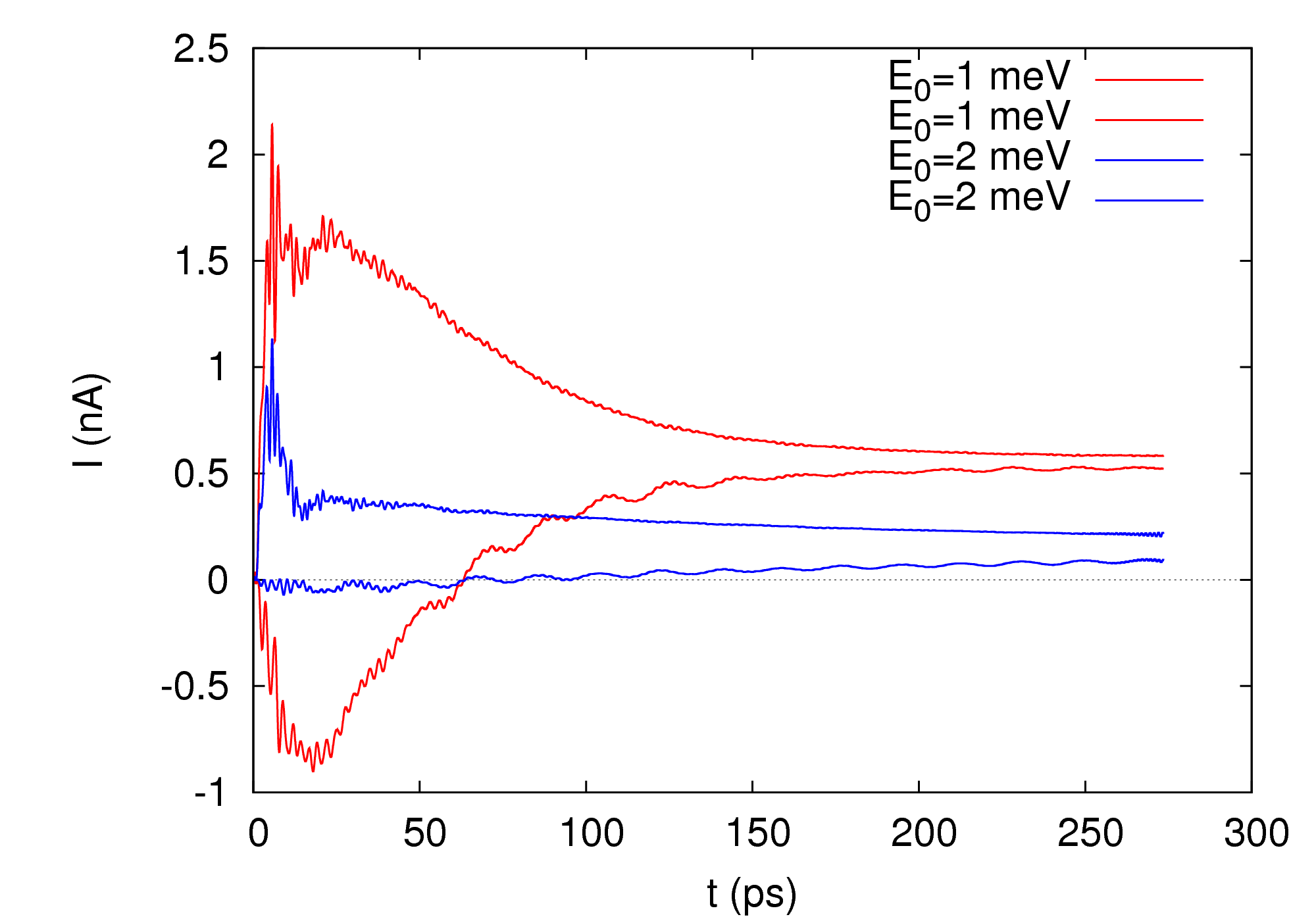}
      \end{center}
      \caption{Total current in the right and left leads for a narrow and a
               broad confinement for $\mu_R=0.1$ meV (upper), and
               $\mu_R=1.1$ meV (lower). A negative current in the right leads
               is a current into the system from the right lead.}
      \label{TotalCurrent}
\end{figure}

We assume that the system is a finite parabolically confined quantum wire of length 300 nm,
and use GaAs parameters, $m^*=0.067m_e$.
The characteristic confinement energy is $\hbar\Omega_0=1.0$ meV.
The ends of the system have a hard wall for $t<0$ and are for later times 
tunnel coupled to the semi-infinite
parabolic leads that will either be broad, with confinement energy scale
$E_0=\hbar\Omega^{L,R}_0=1.0$ meV, or more narrow with $E_0=2.0$ meV.
The energy spectrum of the lowest MESs of the system is seen in Fig.\ \ref{Spectra}
together with the continuous SESs spectra for the broad and the more narrow leads.
The chemical potential of the right lead is fixed at $\mu_R=1.4$ meV and the bias $\Delta\mu$
is varied by changing $\mu_L$. The total time-dependent occupation of the system with electrons
is displayed in Fig.\ \ref{TotalCharge} for the two types of leads and different values of the bias.
The largest amount of charge is accumulated in the system for the higher bias and the
broad leads, and for the broad leads and the lower bias the system is very close to a
steady state regime after the transient charging.
Furthermore, we see in Fig.\ \ref{TotalCurrent} how the charging for the broad
leads occurs through both leads in the transient regime, and how the steady
state value of the current through the system depends on the bias.

\begin{figure}[htbq]
      \begin{center}
      \includegraphics[width=0.42\textwidth,angle=0]{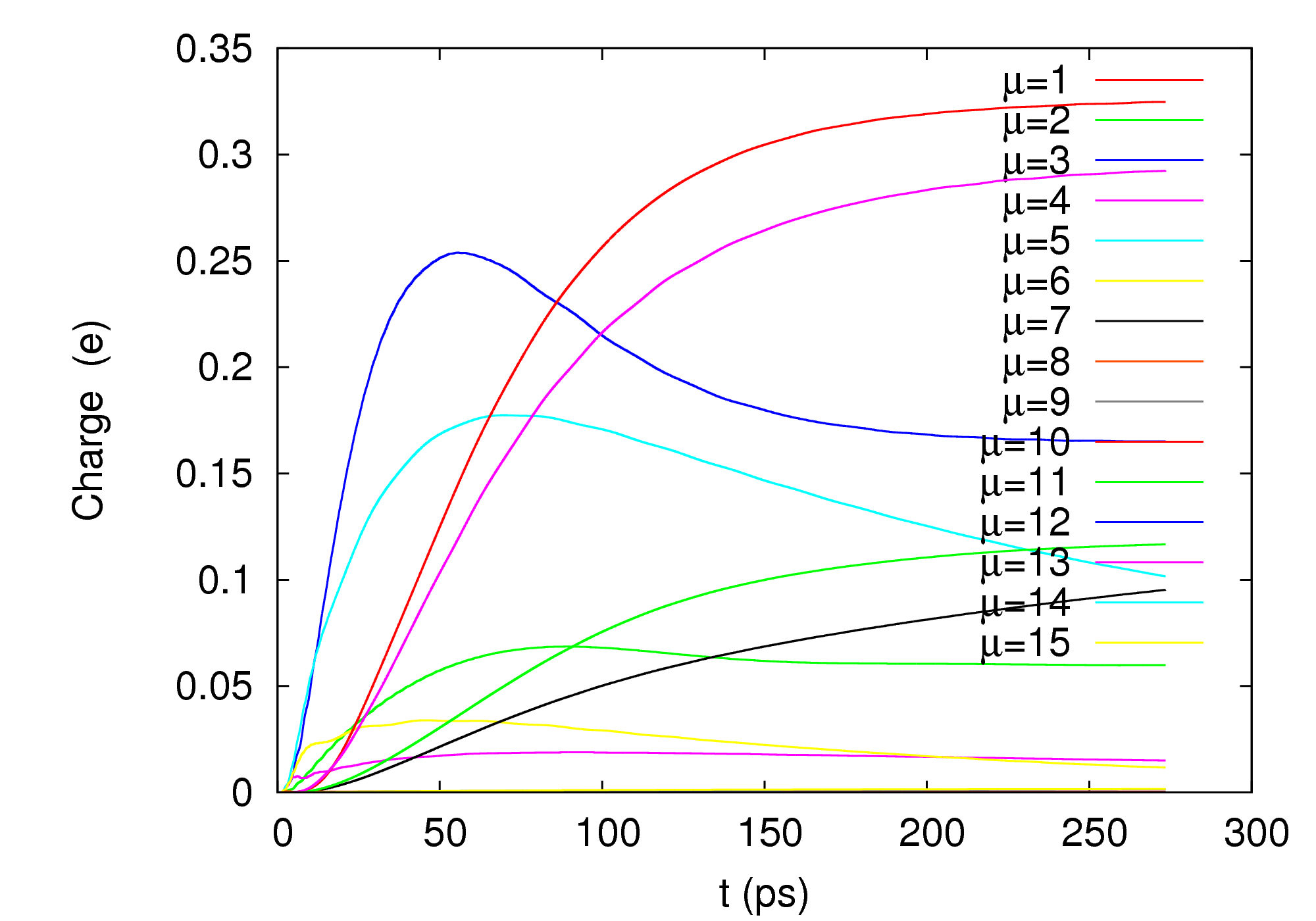}
      \includegraphics[width=0.42\textwidth,angle=0]{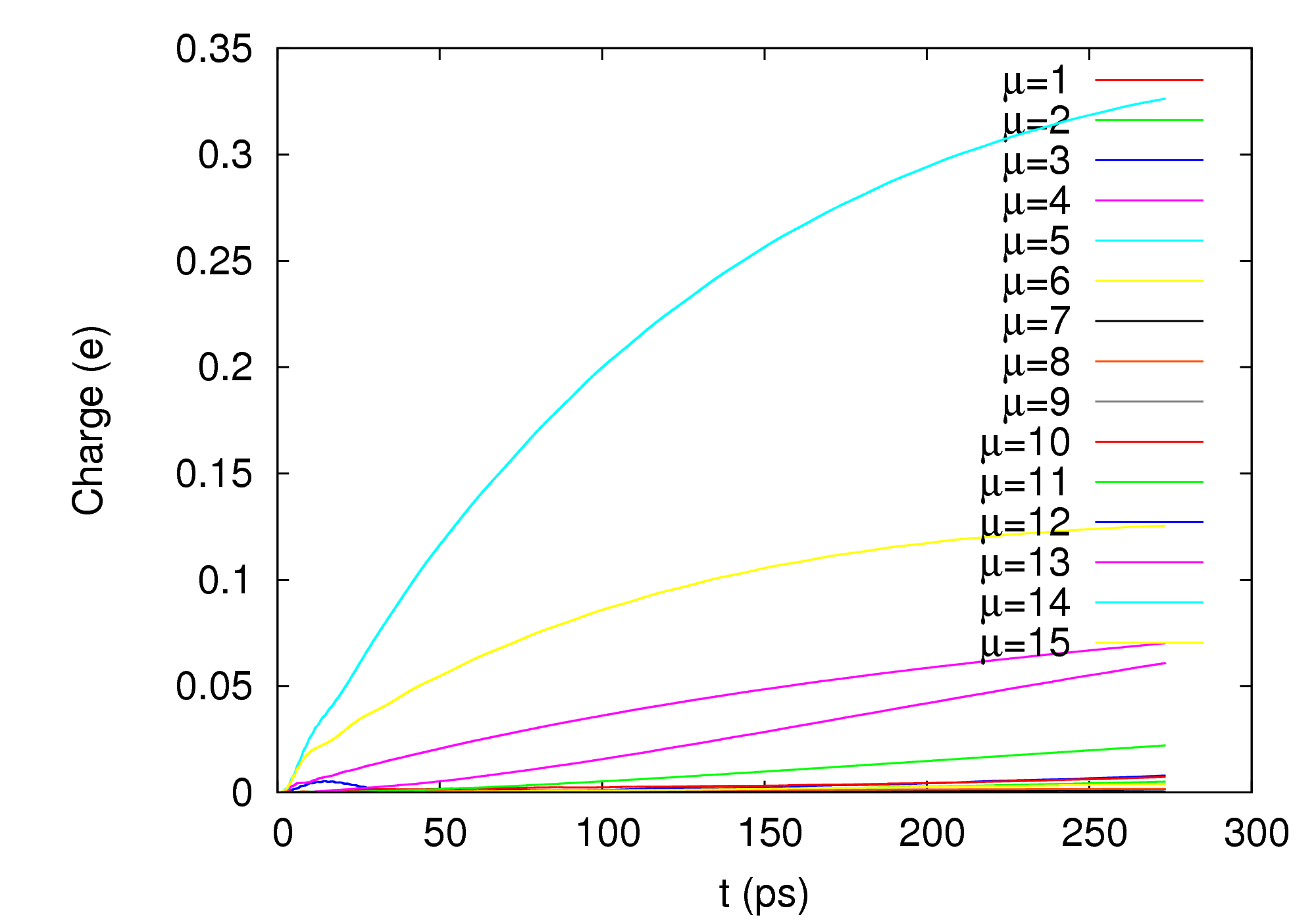}
      \end{center}
      \caption{Partial time-dependent charge in some MESs states for
               $E_0=\hbar\Omega^{R,L}_0=1.0$ meV (upper), and
               $E_0=2.0$ meV (lower). $\Delta\mu =1.1$ meV.}
      \label{PartCharge}
\end{figure}

Interesting is to see how the different MESs contribute to the transport
in Fig.\ \ref{PartCharge}. Not only more MESs participate in the case of the
broad leads, but more interesting is to see their location with respect to $\mu_R$.
If we analyze the structure of the MESs using the Fock space of SESs 
$|\mu\rangle = |i^\mu_1,i^\mu_2,i^\mu_3,\cdots,i^\mu_{N_\mathrm{SES}}\rangle$
with $i^\mu_a\in\{0,1\}$ we find that in case of the broad leads a large
contribution comes from $|01.00000000\rangle$, $|01.10000000\rangle$, and
$|01.01000000\rangle$, where we have used a dot to indicate the location of
$\mu_R$. In the case of the more narrow leads we find large contributions from
$|00.10000000\rangle$, $|00.11000000\rangle$, and  $|00.10100000\rangle$.
In both cases the lowest SES is not occupied since it is a bulk state with 
low coupling to the contact region. For the broad wire we see the state just
under $\mu_R$ is occupied, and states in the bias window. For the narrow
leads this state is not occupied, but rather the next state just above $\mu_R$.
In Fig.\ \ref{Spectra} we see that this can be explained by the lack of low 
enough energy states in the narrow leads. In the case of the
broad leads we notice that initially one-electron states are occupied, but
for later times two-electron states are favored.  

We see thus that the energy spectrum of the leads is very important when
considering which states in the system will contribute to the transport,
both in the transient and the steady state regime. In addition, the character
of the state in the system both with respect to energy and geometry is
essential as the coupling depends on all these details. The coupling to the 
leads enforces certain correlation on the electrons in the system.
To describe these correlation effects in the weak coupling limit for a low
density of electrons we needed the GME many-electron formalism. 
Elsewhere, we have shown that the Coulomb interaction between the
electrons influences the charging of the system by changing the 
energy scales, the spectra, and the correlation between the electron states
\cite{Moldoveanu10:IGME}.   

\subsubsection*{Acknowledgments}
      The authors acknowledge financial support from the Research
      and Instruments Funds of the Icelandic State,
      the Research Fund of the University of Iceland, the
      Icelandic Science and Technology Research Programme for
      Postgenomic Biomedicine, Nanoscience and Nanotechnology, the
      National Science Council of Taiwan under contract
      No.\ NSC97-2112-M-239-003-MY3, and the Reykjavik University Development
      Fund T09001.

\bibliographystyle{elsarticle-num}

\begin{thebibliography}{1}                                     
\expandafter\ifx\csname url\endcsname\relax                    
  \def\url#1{\texttt{#1}}\fi                                   
\expandafter\ifx\csname urlprefix\endcsname\relax\def\urlprefix{URL }\fi
\expandafter\ifx\csname href\endcsname\relax                            
  \def\href#1#2{#2} \def\path#1{#1}\fi                                  

\bibitem{Harbola06:235309}
U.~Harbola, M.~Esposito, S.~Mukamel,
  \href{http://link.aps.org/abstract/PRB/v74/e235309}{Quantum master equation
  for electron transport through quantum dots and single molecules}, Phys. Rev.
  B 74 (2006) 235309.                                                          

\bibitem{Welack06:044712}
S.~Welack, M.~Schreiber, U.~Kleinekath{\"o}fer,
  \href{http://link.aip.org/link/?JCPSA6/124/044712/1}{The influence of
  ultrafast laser pulses on electron transfer in molecular wires studied by a
  non-markovian density-matrix approach}, J. Chem. Phys. 124 (2006) 044712.

\bibitem{Timm08:195416}
C.~Timm, Phys. Rev. B 77 (2008) 195416.

\bibitem{Vaz08:012012}
E.~Vaz, J.~Kyriakidis, Journal of Physics Conference Series 107 (2008) 012012.

\bibitem{Zwanzig60:1338}
R.~Zwanzig, J. Chem. Phys. 33 (1960) 1338.

\bibitem{Nakajima58:948}
S.~Nakajima, Prog. Theor. Phys. 20 (1958) 948.

\bibitem{Moldoveanu09:073019}
V.~Moldoveanu, A.~Manolescu, V.~Gudmundsson,
  \href{http://stacks.iop.org/1367-2630/11/073019}{Geometrical effects and
  signal delay in time-dependent transport at the nanoscale}, New Journal of
  Physics 11~(7) (2009) 073019.

\bibitem{Gudmundsson09:113007}
V.~Gudmundsson, C.~Gainar, C.-S. Tang, V.~Moldoveanu, A.~Manolescu,
  \href{http://stacks.iop.org/1367-2630/11/113007}{Time-dependent transport via
  the generalized master equation through a finite quantum wire with an
  embedded subsystem}, New Journal of Physics 11~(11) (2009) 113007.

\bibitem{Moldoveanu10:IGME}
V.~Moldoveanu, A.~Manolescu, C.-S. Tang, V.~Gudmundsson, Coulomb interaction
  and transient charging of excited states in open nanosystems,
  arXiv:1001.0047.
\end{thebibliography}

\end{document}